\documentclass[]{aa}

\def\rmit#1{{\it #1}}              
\def\specchar#1{{\sc #1}}

\def\SiI{\mbox{Si\,\specchar{i}}}

\def\eg{\rmit{e.g.}}

\usepackage{booktabs}
\usepackage{graphicx}
\usepackage{multirow}
\usepackage{color}
\usepackage[flushleft]{threeparttable}

\usepackage{array}
\newcolumntype{?}{@{\vrule width 2pt}}

\usepackage{hyperref}
\hypersetup{
    final=true,
    pageanchor=true,
    colorlinks=true,
    breaklinks=true,
    linkcolor=blue,
    citecolor=blue,
    urlcolor=blue,
    pdfpagemode=UseNone,
    pdftitle={},
    pdfauthor={T. Felipe et al.},
    pdfsubject={Solar Physics},
    pdfkeywords={Methods: observational -- Sun: photosphere -- Sun: chromosphere -- Sun: oscillations  -- sunspots -- Techniques: polarimetric}}

\titlerunning{Cutoff frequency in solar models}   

\begin{document}



\title{Numerical determination of the cutoff frequency in solar models}

\author{T. Felipe\inst{\ref{inst1},\ref{inst2}}
\and C. R. Sangeetha\inst{\ref{inst1},\ref{inst2}}
}


\institute{Instituto de Astrof\'{\i}sica de Canarias, 38205, C/ V\'{\i}a L{\'a}ctea, s/n, La Laguna, Tenerife, Spain\label{inst1}
\and 
Departamento de Astrof\'{\i}sica, Universidad de La Laguna, 38205, La Laguna, Tenerife, Spain\label{inst2} 
}

\abstract
{In stratified atmospheres, acoustic waves can only propagate if their frequency is above the cutoff value. The determination of the cutoff frequency is fundamental for several topics in solar physics, such as evaluating the contribution of those waves to the chromospheric heating or the application of seismic techniques. However, different theories provide different cutoff values.} 
{We developed an alternative method to derive the cutoff frequency in several standard solar models, including various quiet-Sun and umbral atmospheres. The effects of magnetic field and radiative losses on the cutoff are examined.}
{We performed numerical simulations of wave propagation in the solar atmosphere using the code MANCHA. The cutoff frequency is determined from the inspection of phase difference spectra computed between the velocity signal at two atmospheric heights. The process is performed by choosing pairs of heights across all the layers between the photosphere and the chromosphere, to derive the vertical stratification of the cutoff in the solar models. }
{The cutoff frequency predicted by the theoretical calculations departs significantly from the measurements obtained from the numerical simulations. In quiet-Sun atmospheres, the cutoff shows a strong dependence on the magnetic field for adiabatic wave propagation. When radiative losses are taken into account, the cutoff frequency is greatly reduced and the variation of the cutoff with the strength of the magnetic field is lower. The effect of the radiative losses in the cutoff is necessary to understand recent quiet-Sun and sunspot observations. In the presence of inclined magnetic fields, our numerical calculations confirm the reduction of the cutoff frequency due to the reduced gravity experienced by waves propagating along field lines. An additional reduction is also found in regions with significant changes in the temperature, due to the lower temperature gradient along the path of field-guided waves.}
{Our results show solid evidences of the stratification of the cutoff frequency in the solar atmosphere. The cutoff values are not correctly captured by theoretical estimates. In addition, most of the widely-used analytical cutoff formulae neglect the impact of magnetic fields and radiative losses, whose role is critical to determine the evanescent or propagating nature of the waves. }


\maketitle


\section{Introduction}

The presence of acoustic waves in the solar atmosphere and interior is a well-established fact, and their study is a fundamental subject in solar physics. The cutoff frequency of those waves plays an important role in their propagation in the solar atmosphere given that the Sun is a stratified medium. Acoustic waves whose frequency is above the cutoff frequency can pass through the photosphere and propagate into higher atmospheric layers. On the contrary, those waves with frequency lower than the cutoff value are called evanescent waves and are trapped inside the solar interior. The acoustic cutoff has significant implications for several topics in solar physics. Acoustic waves are one of the candidates proposed for explaining the heating of the outer atmospheric layers \citep{1990ASPC....9....3U,1996SSRv...75..453N, 2003ASPC..286..363U}, and the cutoff frequency determines which waves contribute to the energy flux. The cutoff leads to the existence of global p-mode oscillations produced by waves trapped below the solar surface and, thus, it enables the field of helioseismology \citep{Christensen-Dalsgaard2002}. It is also a key parameter to understand the shift in the dominant period in some solar magnetic structures from 5 minutes at the photosphere to 3 minutes at the chromosphere \citep{Fleck+Schmitz1991, Centeno+etal2006, Felipe+etal2010b} and the more puzzling propagation of long-period waves at chromospheric layers \citep[\eg,][]{Orrall1966, Giovanelli+etal1978}. Observations of these long-period waves, which are expected to be evanescent in the solar atmosphere, has intrigued the interest in them \citep{2007SoPh..246...53D,Jefferies+etal2006,Centeno+etal2009}  

The acoustic cutoff frequency is dependent on the local environment. In the
literature, there are many equations to compute the cutoff frequency. \citet{Lamb1909} carried the initial study to derive the cutoff frequency for an
isothermal atmosphere. Many studies have been carried out for non-isothermal media
\citep[\eg,][]{1984ARA&A..22..593D,2006PhRvE..73c6612M} as the
Sun has a stratified atmosphere. However, it has been shown that the choice of different independent and dependent parameters can generate different analytical equations for the cutoff frequency \citep{Schmitz+Fleck1998,Schmitz+Fleck2003}. The above studies do not incorporate the effects of
the magnetic field. There have been some attempts to evaluate the impact of the magnetic field in the cutoff frequency of the so-called magnetoacoustic waves
\citep[\eg,][]{1982ApJ...262..760T,1983AnRFM..15..321T,Roberts1983,1993ApJ...409..450S,
Roberts2006,2015MNRAS.450.3169P}. Theoretical considerations indicate that, in the low-$\beta$ regime, where $\beta=c_{\rm s}^2/v_{\rm A}^2$, the presence of a magnetic field inclined from the solar vertical an angle $\theta$ produces a reduction of the cutoff frequency by a factor $\cos\theta$ \citep{Bel+Leroy1977, Jefferies+etal2006}. This effect, commonly referred to as ``ramp effect'' or ``magnetoacoustic portals'', is due to the field-guided propagation of the slow mode waves, which experience a reduced gravity along field lines. The reduction of the cutoff frequency of magnetoacoustic waves in atmospheres with inclined magnetic fields has been observationally confirmed  \citep{Jefferies+etal2006,McIntosh+Jefferies2006,Rajaguru+etal2019}.

Comparative studies of the height dependence of the cutoff in solar observations are sparse, although several works confirm the stratification of the cutoff in the solar atmosphere \citep{Wisniewska+etal2016,Felipe+etal2018b}. They also point to significant disagreements between the theoretically computed cutoff values and those actually observed. In this paper we follow a numerical approach to determined the cutoff frequency in several solar models. The results are compared with the cutoff values derived from analytical calculations. We have also compared our simulation results with recent observational evidences of low cutoff frequency at the photosphere
\citep{Rajaguru+etal2019}. The paper is structured as follows. The
numerical methods, solar models, and analytical equations of the cutoff frequency
are discussed in Sect. \ref{sect:methods}. The cutoff frequency derived for various quiet-Sun and sunspot models is discussed in Sect. \ref{sect:cutoff_QS} and \ref{sect:cutoff_umbra},
respectively. The results are discussed and compared with recent observations in Sect. \ref{sect:discussion}. Finally, the conclusions are stated in Sect. \ref{sect:conclusions}.

\section{Methods}
\label{sect:methods}

We have performed numerical simulations of wave propagation in a set of well-known and broadly used standard one-dimensional solar models. The outcome of the simulations has been employed to determine the vertical stratification of the cutoff frequency for each of the models, including a parametric study of the dependence of the cutoff with some parameters of the models, such as the magnetic field strength and the presence of radiative losses. The numerically determined cutoff stratifications have been compared with the cutoff values derived from various formulae described in the literature. In the following sections, we describe the procedures carried out for each of these steps.

\subsection{Numerical simulations}

Numerical simulations in a two-dimensional (2D) domain have been developed using the code MANCHA \citep{Khomenko+Collados2006, Felipe+etal2010a} restricted to the magnetohydrodynamic approximation. The code computes the evolution of the perturbed variables, which are retrieved after explicitly subtracting the background state from the equations. As a background, we have imposed for each simulation a different solar atmosphere. In the vertical direction we have set the stratification of the modified solar models (see next section), whereas the backgrounds are constant in the horizontal direction. Periodic boundary conditions have been imposed in the horizontal direction, and at the top and bottom boundaries we have set a Perfect matched layer \citep[PML,][]{Berenger1994} to damp the waves and avoid undesired reflections.

\begin{figure*}[!ht] 
 \centering
 \includegraphics[width=18cm]{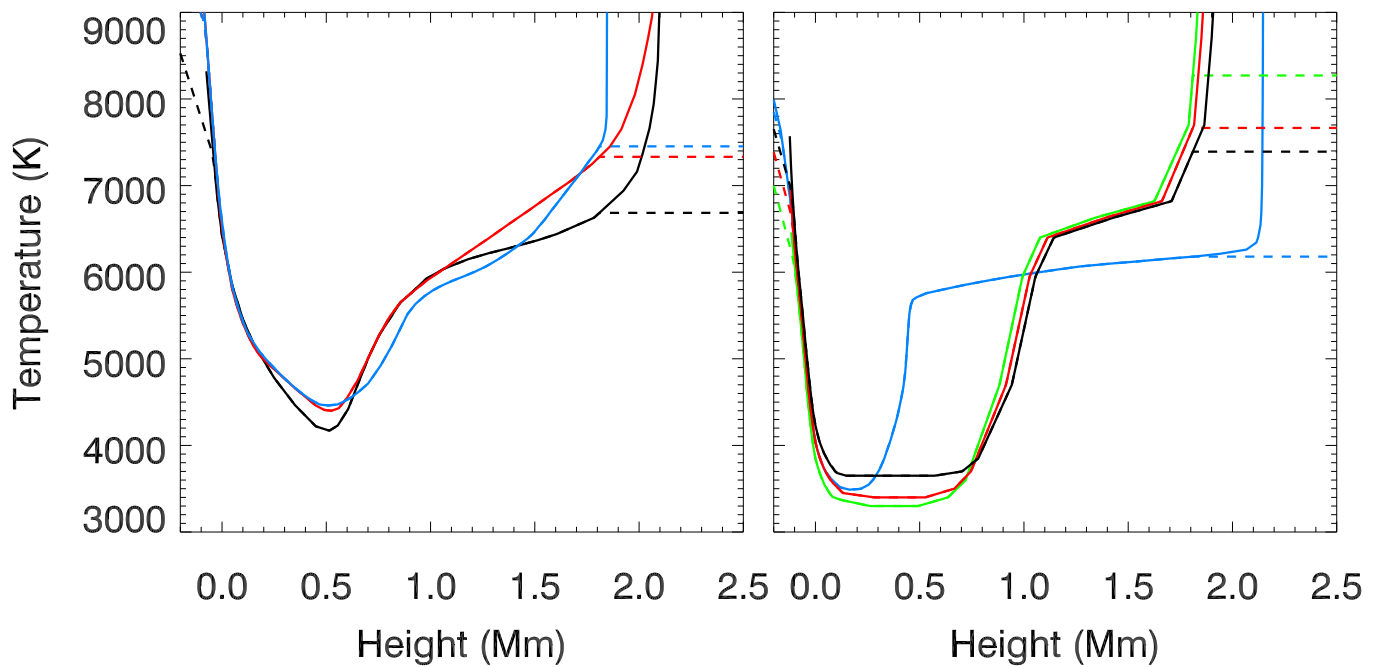}
  \caption{Temperature stratification of the solar models analyzed in this work. Left panel: Quiet-Sun temperature distribution as given by VALC (black line), FALC (red line), and Avrett2015QS (blue line) models. Right panels: Sunspot temperature stratification from eMaltby (green line), mMaltby (red line), lMaltby (black line), and Avrett2015spot (blue line) atmospheres. In both panels, solid lines represent the original models and dashed lines correspond to the temperature profiles employed for the computation of the numerical simulations.}      
  \label{fig:models}
\end{figure*}

We have employed a 2.5D approximation, which means that we use a 2D domain (in the plane $X-Z$, derivatives are not computed in the $y$ direction) but the vectors maintain the three spatial coordinates. In the vertical direction, the computational domain spans from $z=-1,140$ km to $z=2,420$ km, with $z=0$ defined at the height where the optical depth at 5000 \AA\ is unity, and using a constant spatial resolution of 10 km. The PML medium is established in the top ten grid points of the domain. The horizontal domain covers 4,800 km with a spatial step of 50 km. The size of the numerical grid is $96\times356$.

Waves are excited by a vertical force added to the equations. The temporal evolution and spatial dependence of the driving force were taken from temporal series of photospheric Doppler velocity in the \SiI\ 10827 \AA\ line reported in \citet{Felipe+etal2018b}. The details on how the driver is introduced can be found in \citet{Felipe+etal2011}, although in this work we have performed some variations. Here, we are not interested in a one-to-one reproduction of the observed velocities. Instead, we only need to introduce waves that propagate from below the photosphere to the higher chromospheric layers whose properties qualitatively match those found in actual sunspot observations. The driver has been imposed below the photosphere ($z=-180$ km), around 500 km below the formation of the \SiI\ 10827 \AA\ line \citep{Bard+Carlsson2008, Felipe+etal2018b} where the oscillations were measured. In addition, the response of the atmosphere to an oscillatory force depends on the frequency of the oscillations \citep{Felipe+etal2011}. To generate waves with a realistic photospheric power distribution, the amplitude of the power spectra of the driving force has been modified concerning the reference observational velocity. The amplitude of the driver in all the simulations presented in this work is low enough to keep the computations in the linear regime. The duration of each simulation is 65 min of solar time.

\subsection{Solar models}

A total of seven different solar atmospheres have been analyzed, including three quiet-Sun models and four sunspot umbral models. The quiet-Sun atmospheres correspond to those published in \citet{Vernazza+etal1981}, \citet{Fontenla+etal1993}, and \citet{Avrett+etal2015}. In the following, we will refer to them as VALC, FALC, and Avrett2015QS, respectively. In the case of umbral atmospheres, we have explored the three models presented by \citet{Maltby+etal1986}, which correspond to the darkest part of large sunspots at the early (eMaltby), middle (mMaltby), and late (lMaltby) phases of the solar cycle, and the umbral model introduced by \citet{Avrett+etal2015} (Avrett2015spot).          

All of these solar models extend from the photosphere through the chromosphere into the transition region. In this work, we are interested in determining the vertical stratification of the cutoff frequency. At each atmospheric height, we aim to measure the lower frequency where waves begin to propagate upwards. The strong temperature gradients of the transition region reflect the upward propagating waves, and theory predicts that a resonant cavity can be formed between that height and the temperature minimum \citep{Zhugzhda+Locans1981, Zhugzhda2008, Botha+etal2011, Felipe2019, Jess+etal2019}. For our numerical simulations, we have partially removed the top transition region. The temperature has been set constant above $z=1,800$ km for all the solar models. Then, the pressure and density distributions above that height have been computed for the modified temperature stratification following \citet{Santamaria+etal2015}. With this approach, we avoid most of the downward propagating waves coming from the reflections at the transition region, which may affect the estimation of the propagating/non-propagating nature of waves a certain frequencies.

The solar models have also been extended to deeper layers since the bottom boundary of the computational domain employed by our simulations is below the lower limit of these atmospheric models. Below $z=0$, the temperature of the models has been smoothly joined with the temperature stratification of the solar interior from Model S \citep{Christensen-Dalsgaard+etal1996}. We take a constant adiabatic index $\gamma=5/3$.

\subsection{Estimation of the cutoff frequency stratification}

\subsubsection{Determination of the cutoff frequency in numerical simulations}
\label{sect:cutoff_num}

The cutoff frequency has been measured from the examination of the phase difference ($\Delta\phi$) spectra between the vertical velocity at two heights. Low-frequency waves, whose frequency is below the cutoff value, form evanescent waves. In this situation, the signal from the two atmospheric heights oscillate in phase and the phase spectrum exhibits values near zero. From a certain frequency value onward, the phase difference (obtained from the subtraction of the phase of the velocity signal at the lower height from the phase at the higher height) shows a progressive increasing tendency, indicating that these frequency modes propagate from deeper to higher layers. The starting point of the increasing trend corresponds to the cutoff frequency.  

\begin{figure}[!ht] 
 \centering
 \includegraphics[width=9cm]{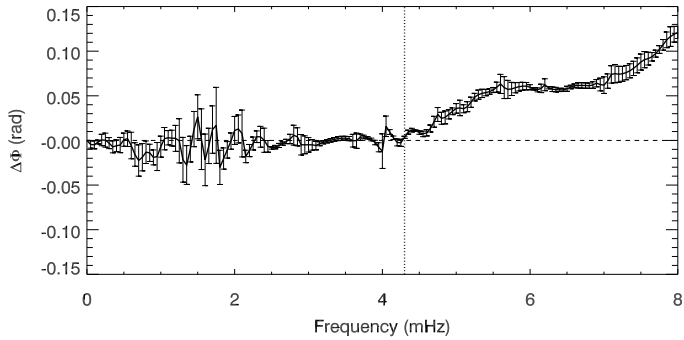}
  \caption{Average phase difference spectra between the vertical velocity signal at $z=880$ km and $z=900$ km measured from the simulation of VALC model with a vertical magnetic field of 130 G. A positive phase difference indicates upward wave propagation. Error bars show the standard deviation of the averaged data. The vertical dotted line marks the value of the cutoff frequency as determined from the examination of the phase spectra.}      
  \label{fig:phase_dif}
\end{figure}

Figure \ref{fig:phase_dif} illustrates the phase difference between two atmospheric heights in one of the numerical simulations performed using VALC model as a background. In this example, the lower height is $z=880$ km, whereas the upper layer corresponds to $z=900$ km. That is, the height difference between both velocity signals is $\Delta h=20$ km. The solid line in Fig. \ref{fig:phase_dif} shows the horizontal average of the phase difference spectra and the vertical bars are the standard deviation of those points. The temporal series of 65 min of simulation have been padded with zeros up to 341 min. According to this data, we have estimated that the cutoff frequency of this solar model at $z=890$ km is 4.30 mHz (indicated by a vertical dotted line). This is the frequency value where the increasing trend of the phase difference starts and where we can assure that there is upward wave propagation, since the phase difference (taking into account the uncertainty estimated from the standard deviation) is undoubtedly positive.

We have computed the cutoff frequency for all the models presented in the previous sections, including various values of the magnetic field, at all heights between $z=0$ and $z=1,800$ km. In all cases, we have employed a $\Delta h=20$ km, and the obtained cutoff value has been assigned to the mid-point between the two heights used for the analysis. A script has been written to automatically determine the cutoff values. To adapt it to provide a good performance despite the peculiarities of each case, the procedures performed by this script slightly depart from the description of the cutoff estimation discussed in the previous paragraph. First, the mean phase difference spectrum has been smoothed by averaging all the phase differences in a box with a width of 0.5 mHz. Then, we have determined the region where the smoothed phase difference, after adding to it the standard deviation, is above zero. These are the frequencies where we consider that waves are propagating. Finally, the lowest frequency of that region is selected as the cutoff value.

\subsubsection{Analytical formulae for the cutoff frequency}

The cutoff frequency is a local quantity which depends on the properties of the surrounding medium. The solar atmosphere exhibits significant inhomogenities, and one would expect changes in the cutoff value both in the horizontal direction as in the vertical direction. In this work, we focus on the stratification of the solar atmospheric models, since the background model from our simulations are constant in the horizontal axis. Theory predicts a wide variety of cutoff formulae depending on the wave equation derived from different selections of independent and dependent variables \citep{Schmitz+Fleck2003}.

The main goal of this work is to bypass this limitation in the derivation of the cutoff frequency by providing alternative values for the cutoff in the solar atmosphere from the examination of numerical simulations. We have also computed the theoretical cutoff stratification in the analyzed models by applying several expressions commonly used in the literature. This allows us to compare the analytical cutoff frequency with the numerical estimations, and discriminate between different theories. 

\begin{figure*}[!ht] 
 \centering
 \includegraphics[width=18cm]{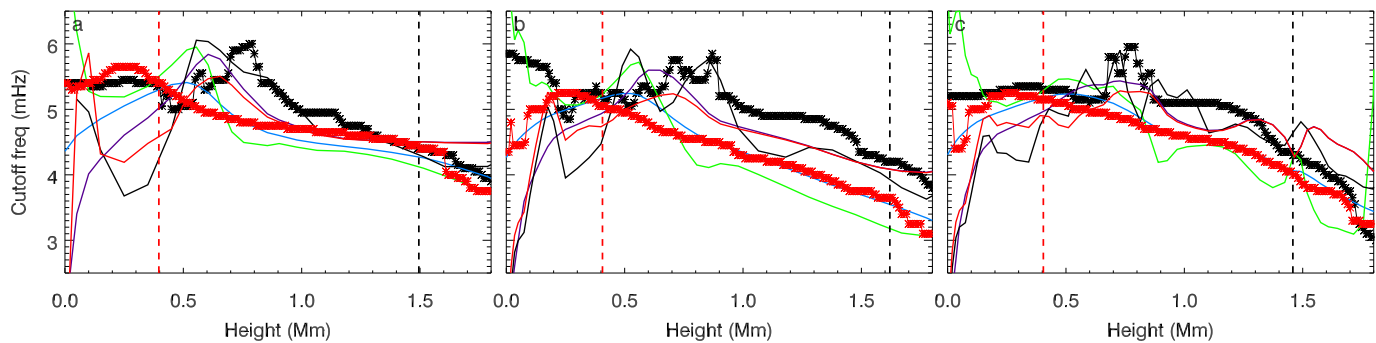}
  \caption{Variation of the cutoff frequency with height in the quiet Sun models VALC (panel a), FALC (panel b), and Avrett2015QS (panel c). The lines with asterisks show the cutoff values determined from the examination of the phase difference spectra in a numerical simulation with a vertical magnetic field of 5 G (black) and 300 G (red). Color lines indicate the analytical cutoff frequency computed using Eq. \ref{eq:wc1} (blue line), Eq. \ref{eq:wc2} (green line), Eq. \ref{eq:wc3} (violet line), Eq. \ref{eq:wc4} with 5 G magnetic field strength (black line), and Eq. \ref{eq:wc4} with 300 G magnetic field strength (red line). The vertical dashed lines mark the height where the plasma-$\beta$ is unity for the models with a field strength of 5 G (black) and 300 G (red).}      
  \label{fig:cutoff_QS_formulae}
\end{figure*}

We have chosen to examine the same four cutoff formulae whose performance was compared with sunspot observational measurements in \citet{Felipe+etal2018b}. They are described in the following lines. In the case of an isothermal atmosphere, the correct formula for the cutoff is obtained from the original work by \citet{Lamb1909}, and it is given by

\begin{equation}
\omega_{\rm C1}=\frac{c_{\rm s}}{2H_{\rm p}},
\label{eq:wc1}
\end{equation}

\noindent where $H_{\rm p}$ is the pressure scale height and $c_{\rm s}$ is the sound velocity. This expression is only valid for isothermal atmospheres, where it does not change with height. However, in this analysis we have assumed that it is a local quantity, and we have computed its vertical variation according to the temperature stratification of the solar models.

The second formula that we have considered is 

\begin{equation}
\omega_{\rm C2}=\frac{c_{\rm s}}{2H_{\rm \rho}}\Big (1-2\frac{dH_{\rm \rho}}{dz}\Big )^{1/2},
\label{eq:wc2}
\end{equation}

\noindent which is the most commonly used in helioseismology. Here, $H_{\rm \rho}$ is the density scale height. We have also evaluated one of the cutoff formulae from \citet{Schmitz+Fleck1998}, which reads as\footnote{Equation 3 from \citet{Felipe+etal2018b} includes a typo in this expression. Here we show the correct form. The correct formula was employed for the results illustrated in \citet{Felipe+etal2018b}.}

\begin{equation}
\omega_{\rm C3}=\omega_{\rm C1}\Big (1+2\frac{\frac{dc_{\rm s}}{dz}}{\omega_{\rm C1}}\Big )^{1/2}.
\label{eq:wc3}
\end{equation} 

These three cutoff frequencies were derived for purely acoustic waves, neglecting the effect of the magnetic field. They are usually employed in magnetized media since the behavior of slow-mode waves in regions dominated by magnetic pressure (high atmospheric layers or even the photosphere in active regions) is similar to that of acoustic waves. The last formula that we have evaluated is extracted from \citet{Roberts2006}, and it is specifically derived for slow magnetoacoustic waves in isothermal atmospheres permeated by a uniform vertical magnetic field. It is computed as

\begin{equation}
\omega_{\rm C4}=c_{\rm t}\Big [\frac{1}{4H_{\rm p}^2} \Big (\frac{c_{\rm t}}{c_{\rm s}}\Big )^4-\frac{1}{2}\gamma g\frac{\partial}{\partial z}\Big (\frac{c_{\rm t}^2}{c_{\rm s}^4}\Big )+\frac{1}{v_{\rm A}^2}\Big (N^2+\frac{g}{H_{\rm p}}\frac{c_{\rm t}^2}{c_{\rm s}^2}\Big )\Big ]^{1/2} 
\label{eq:wc4}
\end{equation} 

\noindent where $c_{\rm t}=c_{\rm s}v_{\rm A}/\sqrt{c_{\rm s}^2+v_{\rm A}^2}$ is the cusp speed, $N^2$ is the squared Brunt-V\"ais\"al\"a frequency, and $g$ is the gravity.

\section{Cutoff frequency in quiet-Sun models}
\label{sect:cutoff_QS}

Following the methodology described in the previous section, we have derived the stratification of the cutoff frequency in the quiet Sun from the examination of numerical simulations and the application of several formulae described in the literature. The two approaches and the different theoretical cutoff values are compared in Sect. \ref{sect:cutoff_QS_formulae}, whereas a parametric study of the dependence of the numerical cutoff with the magnetic field is presented in Sect. \ref{sect:cutoff_QS_Bz}. In Sect. \ref{sect:cutoff_QS_tr} we discuss the effect of the radiative losses.

We have explored three different quiet-Sun models: VALC, FALC, and Avrett2015QS. In all the numerical simulations the atmosphere is permeated by a constant vertical magnetic field. Some of the quiet-Sun atmospheres imposed as a background for the simulations are convectively unstable. That is, at some heights the square of the Brunt-V\"ais\"al\"a frequency is below zero. However, magnetic fields inhibit convection. We have found that a magnetic field strength as low as 5 G is enough to stabilize the three quiet-Sun models studied in this work. 

The analyses of the simulations have been performed focusing on oscillations in the vertical velocity. Waves are driven below the photosphere, at a depth where the sound speed is much higher than the Alfv\'en speed ($\beta\gg1$). In these layers, the oscillatory signal is produced by fast magnetoacoustic waves, whose properties are similar to that of sound waves. At the height where both characteristic velocities are comparable, mode conversion takes place, and part of the energy of the fast magnetoacoustic wave is converted into fast and slow magnetoacoustic waves in the low-$\beta$ region \citep{Cally2006, Cally2007, Schunker+Cally2006, Khomenko+Collados2006}. There, the target of our analysis is the slow magnetoacoustic mode. It is a field-guided wave which behaves like an acoustic wave. In the case of inclined magnetic fields, mode conversion certainly complicates our approach for determining the cutoff frequency. When measuring the phase difference between two layers in the low-$\beta$ region, we should take into account the horizontal displacement of the wavefront. However, it is not trivial to define the lowest height where waves are field-guided, since wave modes around the $\beta=1$ region exhibit some mixed properties. For simplicity, we have chosen to restrict the analysis of quiet-Sun atmospheres to vertical magnetic fields. In this situation, we can confidently quantify the cutoff frequency from phase difference spectra computed between two heights at the same horizontal position. For an evaluation of the effects of magnetic field inclination on the cutoff, see Sect. \ref{sect:cutoff_umbra_incl}.

\begin{figure*}[!ht] 
 \centering
 \includegraphics[width=18cm]{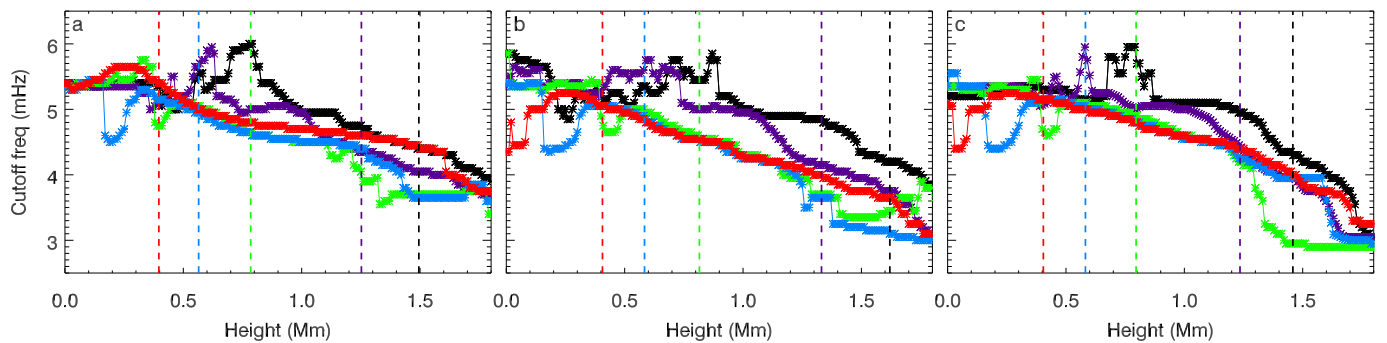}
  \caption{Variation of the numerically determined cutoff frequency with height in the quiet Sun models VALC (panel a), FALC (panel b), and Avrett2015QS (panel c). Each color corresponds to atmospheres permeated by a different strength of the vertical magnetic field: 5 G (black), 10 G (violet), 50 G (green), 130 G (blue), and 300 G (red). The vertical dashed line marks the height where the plasma-$\beta$ is unity following the same color code of the cutoff values. }      
  \label{fig:cutoff_QS_Bz}
\end{figure*}

\subsection{Comparison with analytical models}
\label{sect:cutoff_QS_formulae}

Figure \ref{fig:cutoff_QS_formulae} illustrates the stratification of the cutoff frequency measured in numerical simulations for the VALC, FALC, and Avrett2015QS models, and their comparison with the values obtained from theoretical expressions. The atmospheres are permeated by a vertical magnetic field of $B=5$ G (black lines) or $B=300$ G (red lines). 

In the case with weak magnetic field strength, the cutoff frequency $(\nu_{\rm c})$ of VALC (Fig. \ref{fig:cutoff_QS_formulae}a) shows a peak at around $z=800$ km, where it increases up to ${\sim}$6 mHz. Above that height, the cutoff smoothly decreases, from $\nu_{\rm c}=5.4$ mHz slightly above $z=800$ km to $\nu_{\rm c}\sim 4.0$ at $z=1,800$ km. At the deep photospheric layers, the cutoff exhibits an approximately constant value of ${\sim}$5.4 mHz between $z=0$ and $z=400$ km. It is reduced to $\nu_{\rm c}=5.0$ mHz at $z=500$ km, just before starting to increase to the maximum peak previously mentioned. 

A comparison between the numerically estimated cutoff and the analytical formulae for VALC reveals some qualitative similarities, but significant differences. Strong differences are also found among the four theoretical cutoff estimations. All the analytical stratifications show a maximum peak between $z=500$ km $z=600$ km. This peak is shifted around 200 km towards deeper layers concerning the location of the maximum found in the weakly magnetized simulation. The values of the cutoff at that peak in the analytical expressions are close to the numerically determined value, except for that obtained from Eq. \ref{eq:wc1} for an isothermal atmosphere. At the deep photosphere, the cutoff from Eq. \ref{eq:wc2} shows a reasonable agreement with the numerical cutoff, although in the latter the sharp spike around $z=0$ is missing. At chromospheric heights, the cutoff from Eq. \ref{eq:wc4} provides the best match with the numerical value. However, this expression was derived for slow magnetoacoustic waves and, thus, it is not applicable below the $\beta\sim1$ layer (vertical dashed lines in Fig. \ref{fig:cutoff_QS_formulae}).

The numerical cutoffs obtained for FALC (Fig. \ref{fig:cutoff_QS_formulae}b) and Avrett2015QS (Fig. \ref{fig:cutoff_QS_formulae}c) models share several features with the cutoff of VALC: when the models are permeated by a 5 G magnetic field strength they exhibit a peak around $z=800$ km with a maximum value $\nu_{\rm C}\sim6$ mHz, at chromospheric layers the cutoff value decreases with height, and at the photosphere the cutoff is around 5.3 mHz. The main difference presented by FALC is the high cutoff values of the weakly magnetized case at the deep photosphere, in the bottom 200 km of the plot. This increase agrees with that produced by the cutoff from Eq. \ref{eq:wc2}. Regarding Avrett2015QS model, at chromospheric layers it exhibits a more pronounced reduction of the cutoff with height than the other quiet-Sun atmospheres. Interestingly, in the two models plotted in Fig. \ref{fig:cutoff_QS_formulae}b,c the numerically determined cutoff stratifications for the cases with $B=300$ G exhibit a good agreement with those computed with Eq. \ref{eq:wc1}. The match is almost perfect above $z=800$ km, and the cutoff value at the maximum ($\sim 5.3$ mHz) also agrees, although it is shifted 200 km to deeper layers.

\subsection{Dependence with the magnetic field}
\label{sect:cutoff_QS_Bz}

The solar atmosphere is permeated by magnetic fields even in the regions known as quiet Sun \citep{Trujillo-Bueno+etal2004}. In this section we study the effects of these weak magnetic fields on the quiet-Sun wave propagation by analyzing numerical simulations of VALC, FALC, and Avrett2015QS models with various values of the vertical magnetic field strength. 

Figure \ref{fig:cutoff_QS_Bz} shows the stratification of the cutoff frequency derived from the numerical simulations for the three quiet-Sun models and magnetic field strengths between 5 G (black lines) and 300 G (red lines). In the three atmospheric models the effects of an increasing magnetic field are the same. At most heights, the atmospheres permeated by stronger magnetic fields exhibit lower cutoff frequencies. Some regions depart from this tendency. Deep atmospheric layers are dominated by magnetic pressure, and a lower effect of the magnetic field on the waves is expected at those heights. Surprisingly, in regions with $\beta>1$ the impact of the magnetic field on the derived cutoff frequency is clearly noticeable, and significant differences are found between strongly and weakly magnetized models. 

As the magnetic field increases, the cutoff peak exhibited by all quiet-Sun models progressively decreases and is shifted to deeper layers, from $z\sim800$ km for a strength of 5 G to $z\sim250$ km for a strength of 300 G. In addition, quiet-Sun regions with stronger magnetic fields (above {${\sim}50$ G) also show a more remarkable minimum in the photospheric cutoff, whose location is also shifted to deeper layers as the magnetic field strength increases. In the case of VALC with $B=300$ G, this minimum is out of the range of probed atmospheric heights.

\subsection{Impact of radiative losses on the cutoff}
\label{sect:cutoff_QS_tr}

Radiative losses have been implemented according to Newton's cooling law:

\begin{equation}
Q_{\rm rad}=-c_{\rm v}\frac{T_1}{\tau_{\rm R}}
\label{eq:Qrad}
\end{equation} 

\noindent where $T_1$ is the perturbation in the temperature, $c_{\rm v}$ is the specific heat at constant volume, and the radiative cooling time is given by \citet{Spiegel1957} formula as

\begin{equation}
\tau_{\rm R}=\frac{\rho c_{\rm v}}{16\chi\sigma_{\rm R}T^3}.
\label{eq:spiegel}
\end{equation} 

\noindent In the latter expression, $\rho$ is the density, $T$ is the temperature, $\sigma_{\rm R}$ is the Stefan-Boltzmann constant, and $\chi$ is the gray absorption coefficient. Following \citet{Felipe2019}, the  radiative cooling time given by Eq. \ref{eq:spiegel} has only been applied between $z=200$ km and $z=1,100$ km. \citet{Spiegel1957} formula was derived for optically-thin plasma and assuming local thermodynamic equilibrium. Its range of applicability is restricted to photospheric heights. Out of this region we have imposed adiabatic propagation. Our approach at layers deeper than $z=200$ km is similar to that employed by \citet{Ulmschneider1971}, who assumed a completely optically thick medium (adiabatic propagation) below $z\sim140$ km. At the chromosphere, the effect of the radiative dissipation on wave propagation is negligible \citep{Schmieder1977}. Based on the chromospheric radiative cooling time determined by \citep{Giovanelli1978}, we estimated that the value computed from Eq. \ref{eq:spiegel} is reliable up to $z=1,100$ km, and set adiabatic propagation above that height. A minimum value of $\tau_{\rm R}\sim 10$ s is found at $z=200$ km, and it increases to a maximum of $\tau_{\rm R}\sim 500$ s at $z=700$ km. 

Figure \ref{fig:cutoff_QS_tr} compares the stratification of the cutoff frequency in the quiet-Sun model VALC in the adiabatic case (asterisks) with that measured for simulations where radiative losses are taken into account (diamonds). The results are illustrated for three different values of the magnetic field, including a very quiet atmosphere ($B=5$ G), a model with average quiet-Sun field strength ($B=130$ G), and a case with a stronger magnetic field ($B=300$ G). Radiative transfer is known to reduce the cutoff frequency \citep{Roberts1983, Centeno+etal2006, Khomenko+etal2008b}. Our measurements indicate that this reduction can be striking. When the radiative losses are turned on, the maximum value of the cutoff frequency is $\sim5$ mHz, whereas in the adiabatic cases it shows peaks as high as 6 mHz ($B=5$ G) or 5.7 mHz ($B=300$ G). Radiative losses also lead to low cutoff values around 4 mHz at the photosphere (between $z=150$ km and $z=350$ km) and the low chromosphere ($z\sim1,000$ km). Interestingly, with the radiative losses on, all the simulations show a similar trend in the cutoff stratification, regardless of their magnetic field strength. However, they conserve some dependence with the magnetic field. The relation between the field strength and the cutoff is similar to that found for the adiabatic case, with a general reduction of the cutoff with magnetic field strength for simulations in the range $B=[5, 130]$ G and an increase for $B=300$ G.

A magnetic field permeating the quiet-Sun atmosphere can produce significant variations in the cutoff, including the presence and location of peaks or minimums (see previous section), but those changes are overcome by the effect of radiative transfer.

\begin{figure}[!ht] 
 \centering
 \includegraphics[width=9cm]{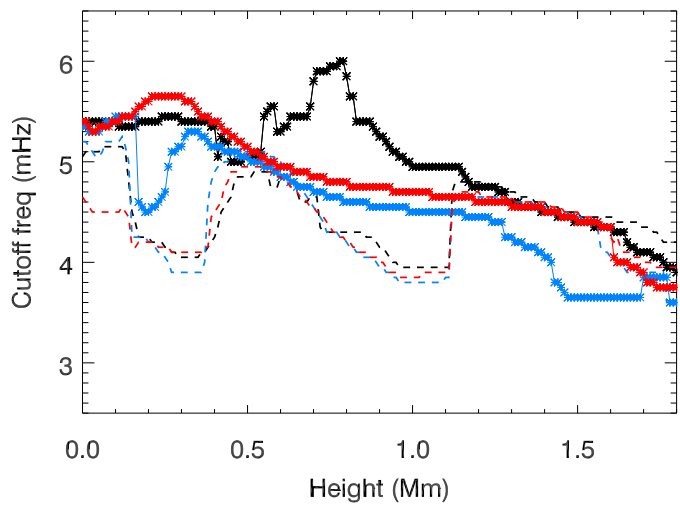}
  \caption{Variation of the numerically determined cutoff frequency with height in quiet Sun models VALC with the radiative looses turned on (dashed lines) and off (solid lines with asterisks). The color indicates the magnetic field strength following the same color code from Fig. \ref{fig:cutoff_QS_Bz}: 5 G (black), 130 G (blue), and 300 G (red).}      
  \label{fig:cutoff_QS_tr}
\end{figure}

\section{Cutoff frequency in umbral models}
\label{sect:cutoff_umbra}

We have evaluated the variation of the cutoff frequency with height in four different umbral models, the three models from \citet{Maltby+etal1986} (eMaltby, mMaltby, lMaltby) and the sunspot model presented by \citet{Avrett+etal2015} (Avrett2015spot). The atmospheres have been permeated by a set of values of a constant vertical magnetic field, from weakly magnetized umbrae (500 G) to strongly magnetized umbrae (3,000 G). We have followed the same approach from the previous section, first comparing the four atmospheres permeated by a chosen value of the magnetic field with the cutoffs derived from the analytical expressions, and then evaluating the variation of the cutoff frequency produced by changes in the magnetic field and radiative losses.

\begin{figure*}[!ht] 
 \centering
 \includegraphics[width=18cm]{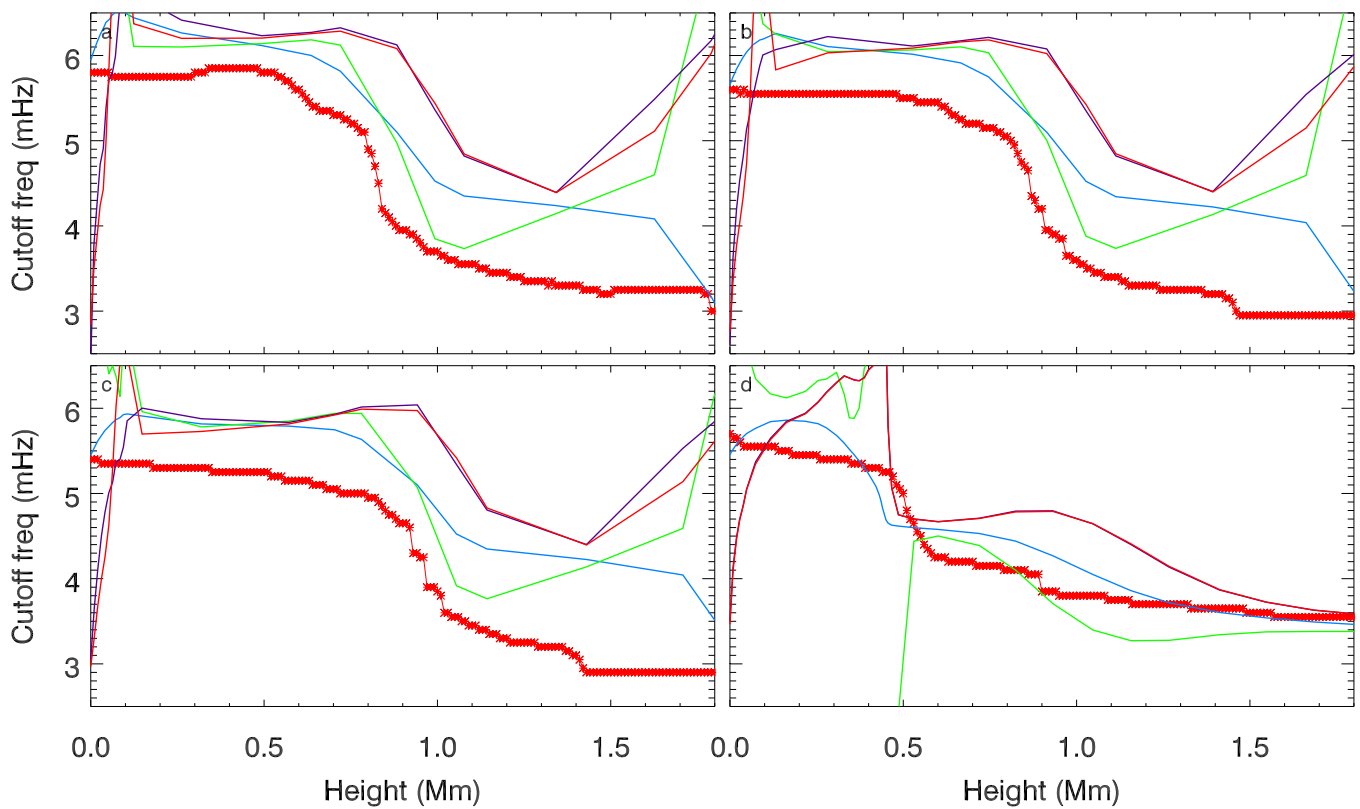}
  \caption{Variation of the cutoff frequency with height in the umbral models eMaltby (panel a), mMaltby (panel b), lMaltby (panel c), and Avrett2015spot (panel d). The red lines with asterisks show the cutoff values determined from the examination of phase difference spectra in numerical simulations with a vertical magnetic field of 3000 G. Color lines indicate the analytical cutoff frequency computed using Eq. \ref{eq:wc1} (blue line), Eq. \ref{eq:wc2} (green line), Eq. \ref{eq:wc3} (violet line), and Eq. \ref{eq:wc4} (red line). All the plotted heights are in the $\beta<1$ region.}      
  \label{fig:cutoff_spot_formulae}
\end{figure*}

\subsection{Comparison with analytical models}

Figure \ref{fig:cutoff_spot_formulae} shows the cutoff frequency of the umbral models with a vertical magnetic field of 3,000 G. As seen in Fig. \ref{fig:models}, the three atmospheres from \citet{Maltby+etal1986} present a similar temperature stratification, but with some small differences. The minimum value of the temperature is related to the intensity of the umbra. Dark umbrae (eMaltby) have lower temperatures, whereas it increases for average (mMaltby) and bright (lMaltby) umbrae. In addition, the height where the temperature starts to rise above the temperature minimum is also shifted to higher layers as brighter umbrae are considered. These differences in the atmospheric stratification are captured by our cutoff measurements. Models with a lower temperature exhibit a higher maximum in the cutoff frequency. In the case of eMaltby, a maximum cutoff of 5.9 mHz is found around $z=400$ km, whereas the maximum cutoff frequencies for models mMaltby and lMaltby are 5.6 mHz and 5.4 mHz, respectively. This dependence of the cutoff with temperature is predicted by theory, and all the analytical expressions explored in this work show a reduction in the photospheric cutoff from eMaltby to lMaltby, with the value of mMaltby located in between. However, all these theories overestimate the cutoff at the umbral photosphere.

The temperature gradients above the temperature minimum (around $z\sim 800$ km) also leave an imprint in the cutoff frequencies. For dark umbrae (eMaltby, Fig. \ref{fig:cutoff_spot_formulae}a) there is a sharp reduction in the cutoff value at $z\sim 800$ km. In the case of mMaltby and lMaltby, the reduction of the cutoff takes place at slightly higher layers, in agreement with the differences presented in the temperature stratification of the models. This sudden cutoff variation is also captured by the theoretical formulae but, again, there are significant differences between the analytically estimated values and those measured in the simulations.   

Regarding the umbral model Avrett2015spot, the general properties of the cutoff stratification are similar to those described for the umbral models from \citet{Maltby+etal1986}. As expected from its temperature profile, the reduction in the cutoff is found at a lower height than in the other models ($z\sim 500$ km). For this atmosphere, the cutoffs from Eqs. \ref{eq:wc2}-\ref{eq:wc4} predict a very high peak at photospheric layers. This peak has not been measured in the numerically determined cutoff. Equation \ref{eq:wc2} gives imaginary cutoff values at $z\sim 450$ km. At the chromosphere, different theories estimate a wide range of cutoff values, and they converge towards the higher layers. The measured cutoff is somewhat in between the predictions from the analytical models.

\begin{figure*}[!ht] 
 \centering
 \includegraphics[width=18cm]{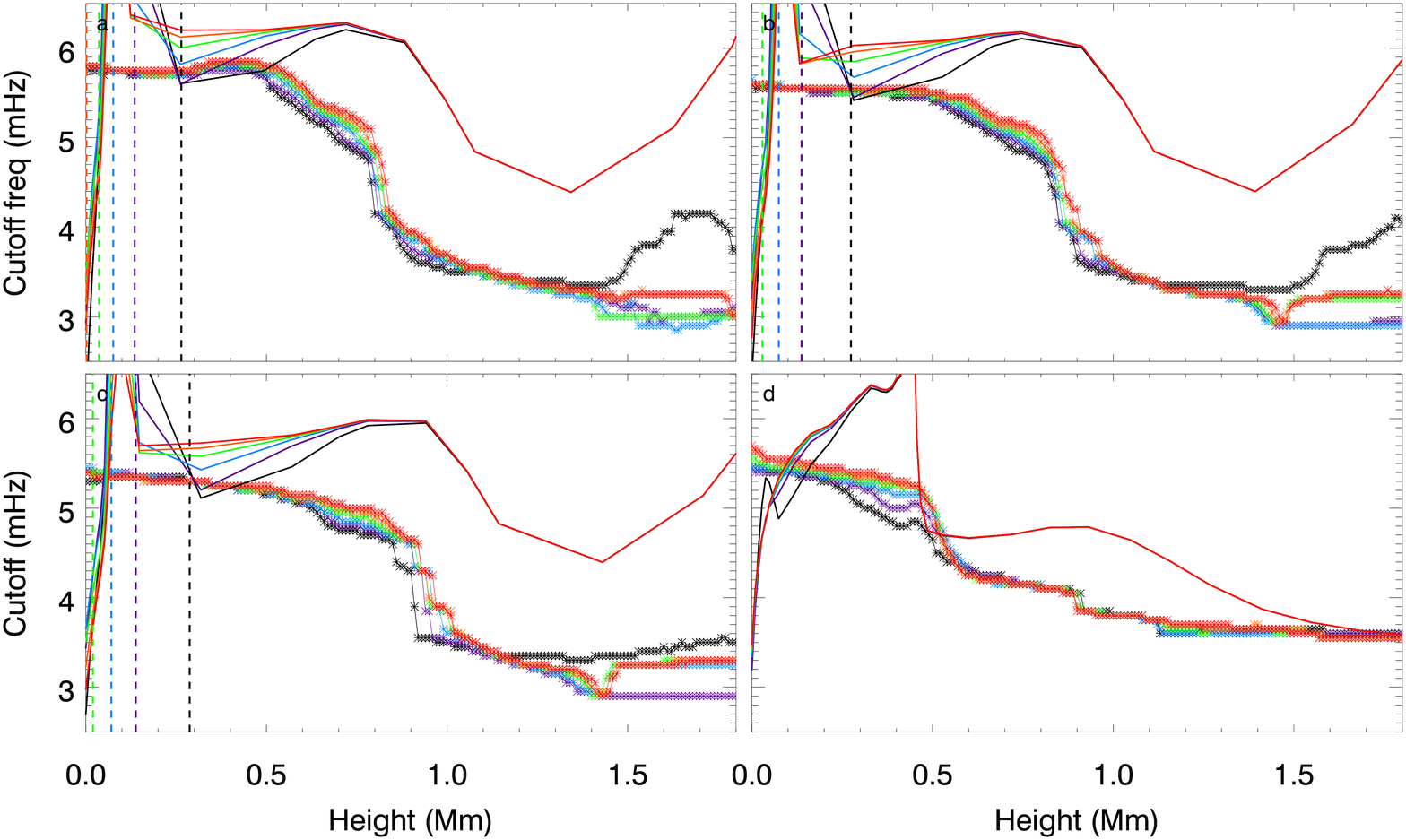}
  \caption{Variation of the numerically determined cutoff frequency with height in the umbral models eMaltby (panel a), mMaltby (panel b), lMaltby (panel c), and Avrett2015spot (panel d). Each color corresponds to atmospheres permeated by a different strength of the vertical magnetic field: 500 G (black), 1000 G (violet), 1500 G (light blue), 2000 G (green), 2500 G (orange), 3000 G (red). The vertical dashed lines mark the height where the plasma-$\beta$ is unity following the same color code of the cutoff values. For some of the atmospheres the line indicating the $\beta=1$ height is not visible because it is below $z=0$. Solid lines show the cutoff frequency of slow magnetoacoustic in atmospheres as given by Eq. \ref{eq:wc4}. Their color indicates the strength of the magnetic field.}      
  \label{fig:cutoff_spot_Bz}
\end{figure*}

\subsection{Dependence with the magnetic field}

The impact of the vertical magnetic field on the umbral cutoff frequency has been evaluated by performing a set of 24 numerical simulations. They correspond to six simulations with different values of the field strength for each of the four umbral models under study. The chosen sample of magnetic field strengths spans from 500 to 3,000 G with a step of \hbox{500 G}. Figure \ref{fig:cutoff_spot_Bz} illustrates the measured cutoff stratifications and their comparison with the theoretical estimates of the slow mode cutoff given by Eq. \ref{eq:wc4}.  

At the deep photosphere, the effect of the magnetic field on the cutoff of the sunspot models is negligible. At those layers the plasma-$\beta$ is around unity. Its value depends mainly on the magnetic field strength. In the case of the \citet{Maltby+etal1986} models, the atmospheres with a magnetic field below 2000 G exhibit a $\beta>1$ in at least some of the analyzed photospheric heights, whereas atmospheres with stronger fields are in the $\beta<1$ regime at all heights above $z=0$. Our results indicate that at those layers where magnetic pressure and gas pressure are comparable, the cutoff frequency does not depend significantly on the magnetic field. In the same way, mode conversion is not relevant for the cutoff, since at a fixed photospheric height strongly magnetized atmospheres (with $\beta<1$, so the vertical velocity signal is a slow mode wave) and weakly magnetized atmospheres ($\beta>1$, the vertical velocity corresponds to oscillations of the fast mode) exhibit similar cutoff values. 

The differences in the cutoff associated with the magnetic field are found at higher layers. As previously discussed, the cutoff in sunspot models decreases from a relatively high photospheric value to lower values at the chromosphere. The location of this variation depends on the atmospheric stratification. In addition, the magnetic field changes the height where this decrease in the cutoff starts. The lower the magnetic field, the deeper the beginning of the cutoff reduction. In \citet{Maltby+etal1986} models, between $z=500$ km and $z=1,000$ km the atmospheres with weaker magnetic field show a lower cutoff frequency. For Avrett2015spot, the region where magnetic field impacts the cutoff is shifted to heights between $z=200$ km and $z=500$ km. This is in contrast with the analytical results from  Eq. \ref{eq:wc4}, where the main differences produced by the magnetic field are found at lower layers, around \hbox{$z=300$ km} for the \citet{Maltby+etal1986} atmospheres and at \hbox{$z=100$ km} for Avrett2015spot. However, the effects of the magnetic field on the cutoff go in the same direction in both numerical and theoretical estimations, with a reduction of the cutoff associated with lower magnetic field strengths.

At the high chromosphere (above $z\sim 1,400$ km), the cutoff frequencies measured for \citet{Maltby+etal1986} models show striking differences between the weakly magnetized umbrae ($B=500$ G, black lines) and the rest of the cases (with magnetic fields stronger or equal to 1000 G). The cases with $B=500$ G exhibit an increase in the cutoff. This is produced by the steep temperature gradient at the beginning of the transition region. In these simulations, only a few hundred kilometers of the transition region are included in \citet{Maltby+etal1986} background models, since above $z=1,800$ km a constant temperature is set (see dashed lines from Fig. \ref{fig:models}b). The simulation using the umbral model eMaltby contains the larger temperature contrast (around 1,400 K temperature difference between $z=1,600$ km and $z=1,800$ km), and it exhibits the larger increase in the chromospheric cutoff (black line from Fig. \ref{fig:cutoff_spot_Bz}a). On the contrary, for lMatlby our simulation only includes a temperature contrast of 600 K, and the chromospheric cutoff increase is more modest (black line from Fig. \ref{fig:cutoff_spot_Bz}c). The presence of magnetic fields stronger than 1,000 G inhibits the cutoff increase associated with those steep gradients at the base of the transition region. The simulation of the Avrett2015spot umbra does not include any steep gradient at the high chromosphere (dashed blue line in Fig. \ref{fig:models}b). In this case, at high layers where the atmosphere is strongly dominated by the magnetic field, the cutoff is similar for all the simulations despite their different field strengths. In this sense, this is in agreement with the predictions of the cutoff from Eq. \ref{eq:wc4}, which do not show dependence on the magnetic field at the chromosphere.

\subsection{Dependence with the field inclination}
\label{sect:cutoff_umbra_incl}

We have chosen to focus on the umbral model mMaltby, since it represents an average umbra and it is probably the most widely-used umbral atmospheric model. For these simulations, we have set a magnetic field strength of 3000 G, in order to assure a $\beta<1$ in the whole umbral atmosphere. We have compared two simulations, one of them with a vertical magnetic field ($\theta=0^{\circ}$) and the other with $\theta=10^{\circ}$. Since we are studying umbral atmospheres, where the magnetic field is mostly vertical, we have restricted the analysis to moderate field inclinations.

The numerical cutoff frequency has been determined following the same procedures described in Sect. \ref{sect:cutoff_num}. In the case with an inclined magnetic field, we have taken into account the horizontal displacement of the waves as they propagate upwards along magnetic field lines. In order to be consistent with the cases with a vertical magnetic field, the distance between the velocity signals used for the computation of the phase spectra has been maintained at \hbox{$\Delta h=20$ km}. However, it is measured along the path of the waves, that is, departing an angle $\theta$ from the vertical. Thus, the horizontal displacement is $\Delta h\sin\theta$, whereas the height difference is $\Delta h\cos\theta$. The velocity field obtained as the output of the simulation has been interpolated to those locations, in order to extract the velocity signal required for the computation of the phase difference spectra.  

Figure \ref{fig:cutoff_spot_incl} illustrates the comparison of the cutoff values determined for the umbral atmosphere with vertical (red) and inclined (blue) magnetic field. The dashed blue line shows the cutoff of the vertical case multiplied by $\cos\theta$. This is the expected cutoff frequency of the inclined case if only the reduced gravity of the ramp effect modifies wave propagation. The dashed blue line agrees with the measured cutoff around $z=500$ km and above $z\sim 1,000$ km. These are the regions where the stratification of the umbral model presents small temperature gradients (red line from Fig. \ref{fig:models}b). The region around $z=500$ km corresponds to the plateau of constant temperature around the temperature minimum, whereas between $z\sim 1,000$ km and the transition region the temperature shows a progressive increase. On the contrary, in those heights where sudden changes in the temperature are found, the measured cutoff is lower than that expected from the ``ramp effect''. This is clearly seen at $z\sim 120$ km and, especially, between $z=700$ km and $z=1,000$ km. Slow mode waves propagate along field lines. Along this path, the temperature contrast that they experience is more gentle than that of vertically propagating waves. This way, the effects of temperature gradients in the cutoff are reduced for waves propagating at a certain angle from the vertical. This reduction in the cutoff is added to that predicted by the reduced gravity along inclined field lines.

\begin{figure}[!ht] 
 \centering
 \includegraphics[width=9cm]{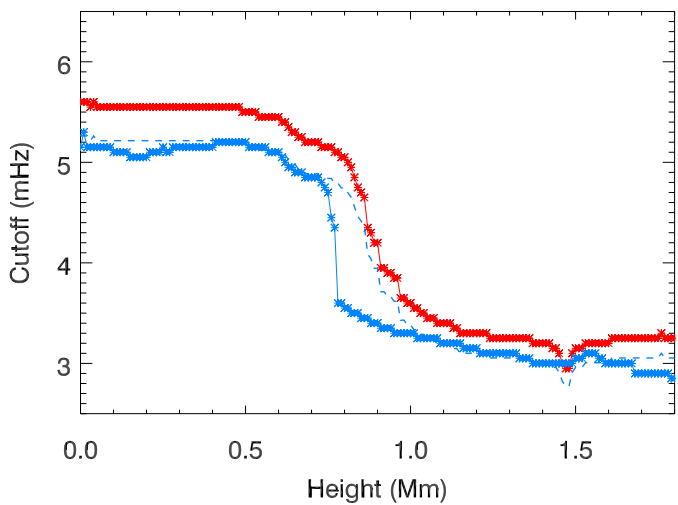}
  \caption{Variation of the numerically determined cutoff frequency with height in the umbral model mMaltby permeated by a 3,000 G magnetic field. Red asterisks show the measurements for vertical magnetic field, and blue asterisks for a field inclination from the vertical of $\theta=10^{\circ}$. Dashed blue line corresponds to the cutoff of the case with vertical magnetic field multiplied by the cosine of the inclination $\theta=10^{\circ}$.}      
  \label{fig:cutoff_spot_incl}
\end{figure}

\subsection{Impact of radiative losses on the cutoff}
\label{sect:cutoff_spot_tr}

Radiative losses have been implemented in an umbral simulation following the same approach described in Sect. \ref{sect:cutoff_QS_tr}. According to Eq. \ref{eq:spiegel}, the radiative cooling time is significantly higher in the umbra than in the quiet Sun, which means that umbral wave propagation is closer to adiabatic propagation. A minimum $\tau_{\rm R}\sim 110$ s is found at $z=200$ km, whereas a maximum value of $\tau_{\rm R}\sim 1800$ s is obtained at $z=900$ km.

Figure \ref{fig:cutoff_spot_tr} shows the measurements of the cutoff for simulations in the adiabatic regime (asterisks) and with radiative losses (diamonds), all of them using the umbral model mMaltby. As an example, the figure illustrates an umbra with a weak magnetic field ($B=500$ G, black) and an umbra with a strong magnetic field ($B=2,500$ G, orange). In both cases the field is vertical.

The introduction of radiative losses produces a reduction of the cutoff in the deep photosphere (around $z=200$ km) and at the base of the chromosphere (between $z\sim 950$ km and $z\sim 1,100$ km, above this height adiabatic propagation is implemented). This is in agreement with the effects found in the quiet-Sun simulations. However, around $z=600$ km the radiative transfer produces a small increase in the cutoff, independently of the magnetic field strength of the umbra. In addition, the increase of the cutoff found in the weakly magnetized umbra associated to the temperature gradients at the base of the transition region is more pronounced in the simulation where radiative losses are turned on, despite that at those regions the wave propagation is adiabatic. The causes of this result are not clear.

\begin{figure}[!ht] 
 \centering
 \includegraphics[width=9cm]{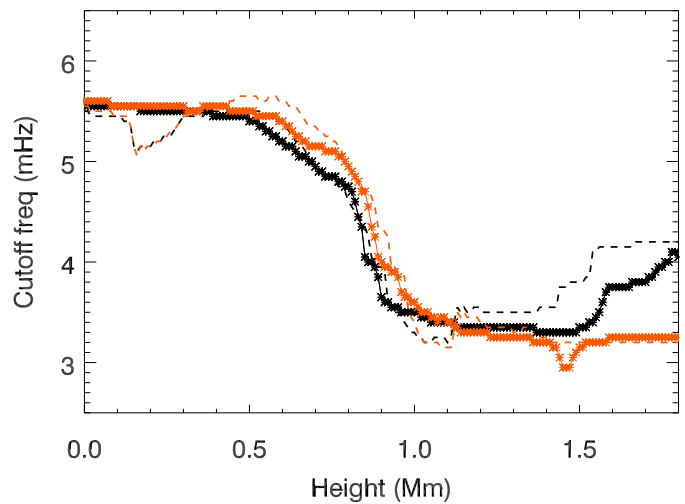}
  \caption{Variation of the numerically determined cutoff frequency with height in umbral model mMaltby with the radiative looses turned on (dashed lines) and off (solid lines with asterisks). Black lines correspond to atmospheres permeated by a 500 G vertical magnetic field, and orange lines to atmospheres with a 2500 G field strength.}      
  \label{fig:cutoff_spot_tr}
\end{figure}





\section{Discussion}
\label{sect:discussion}

The determination of the cutoff frequency is fundamental for understanding wave propagation in atmospheres stratified by gravity. Theoretically, different representations of the wave equation have been derived, and they lead to various forms of the cutoff frequency, which yield significant differences in the cutoff stratification of solar models \citep{Schmitz+Fleck1998}. In this work, we have carried out an alternative approach to determine the cutoff based on the use of numerical simulations. The cutoffs were determined from the analysis of phase difference spectra between the velocity signals measured at two different atmospheric heights. This method has been previously employed in the context of multi-height ground-based observations \citep[\eg,][]{Centeno+etal2006, Wisniewska+etal2016, Felipe+etal2018b} and differs from other works determining the cutoff in numerical simulations from the evaluation of the height-dependent dominant frequency in power spectra \citep{Murawski+etal2016,Murawski+Musielak2016}. A small height difference of $\Delta h=20$ km was chosen between the two heights employed for the computation of the phase spectra, which allows a detailed sampling of the vertical variation of the cutoff. Our quantification of the cutoff is not affected by details in the derivation of the wave equation. In addition, it allows an easy addition of several layers of physics, such as the magnetic field or the radiative losses, which can be difficult to model analytically. We found significant differences between the numerically determined cutoff and those derived theoretically.   

Our results do not clearly favor any of the analytical formulae. Neglecting the effects of magnetic fields and radiative losses, the cutoff measured for the deep photospheric layers of quiet-Sun models is better captured by Eq. \ref{eq:wc2}, which is commonly employed in helioseismic works. However, the sharp spike exhibited by this expression around $z=0$ is not present in the numerical cutoff. Our measurements validate the use of alternative cutoff expressions where the spike is absent. This is the approach followed by several works using the Wentzel, Kramers, and Brilloiuin (WKB) approximation \citep{Cally2007, Moradi+Cally2008}, which is only applicable for wavelengths shorter than the characteristic scales of the medium and, thus, it is inconsistent with the sudden variations in the cutoff given by Eq. \ref{eq:wc2}. We found that in quiet-Sun atmospheres permeated by moderate magnetic field strengths (between 50 and 300 G), the application of the simple formula for isothermal atmospheres (Eq. \ref{eq:wc1}), assuming that it is stratified according to the local temperature, provides the best estimation of the actual cutoff values for adiabatic waves.

Some recent works have measured the stratification of the cutoff in quiet-Sun \citep{Wisniewska+etal2016} and umbral \citep{Felipe+etal2018b} atmospheres based on the analysis of phase difference spectra from multi-height observations employing several spectral lines. \citet{Wisniewska+etal2016} estimated that waves with frequency as low as $\sim 4.2$ mHz can propagate in the quiet-Sun photosphere (around $z=270$ km) and that the cutoff increases up to $\sim 5$ mHz in the next 100 km. According to our findings, this is in qualitative agreement with the propagation of adiabatic waves in an atmosphere permeated by a magnetic field with a specific strength around 130 G (blue lines in Fig. \ref{fig:cutoff_QS_formulae}) or with non-adiabatic wave propagation, regardless of the magnetic field. The second alternative seems more plausible. Regarding umbral atmospheres, \citet{Felipe+etal2018b} found a maximum in the cutoff frequency of 6 mHz around $z\sim500$ km. At deeper photospheric heights it decreases to 5 mHz at $z\sim250$ km. This variation in the umbral photospheric cutoff is not captured by any model with adiabatic propagation (Fig. \ref{fig:cutoff_spot_Bz}). On the contrary, if radiative losses are turned on, the cutoff stratification of the model shows a significant agreement with the observational results (Fig. \ref{fig:cutoff_spot_tr}). Our cutoff estimate also reproduces the chromospheric value slightly above 3 mHz measured by \citet{Felipe+etal2018b}. Both in quiet-Sun and umbral atmospheres, the role of radiative losses is fundamental to understand the cutoff frequency of the observed waves. Note that we have neglected the effect of wave reflections at the transition region. This strategy allows a direct evaluation of the analytical solutions from theoretical models, but challenges the comparison with observational results and, thus, such comparison must be interpreted with care.

Our results are apparently in contrast with the findings from \citet{Heggland+etal2011}. They performed numerical simulations with a sophisticated treatment of radiative losses, and found that they have little effect on the propagation of low-frequency waves. In this work, we have approximated the radiative losses with the Newton's cooling law. This method offers a simple parametrization of the radiative transfer, since it is characterized by the radiative cooling time, and suits our purpose of comparing the numerical cutoffs with those derived from the theory. Our findings agree with the simulations from \citet{Khomenko+etal2008b}, who also employed the Newton's cooling law and found that the energy exchange by radiation can reduce the cutoff frequency. Here, we have employed a variable cooling time as given by \citet{Spiegel1957}. At the photosphere, where this expression is reliable, we obtain a significant reduction in the cutoff. At the high-photosphere/low-chromosphere, the impact of radiative transfer on the cutoff is lower (at those heights the radiative cooling time is higher, so propagation is closer to adiabatic), and waves still need to bypass a cutoff of 5 mHz to reach the chromosphere. This is in agreement with the results from \citet{Heggland+etal2011}. However, many observations have reported long-period waves in regions with vertical magnetic field \citep[\eg,][]{Centeno+etal2009, Stangalini+etal2011, Rajaguru+etal2019}. The reduction of the cutoff produced by radiative transfer is the best candidate to explain those observations.  

Heating by waves is one of the mechanisms proposed to balance the radiative losses of the outer solar atmospheric layers \citep{Biermann1946, Schwarzschild1948}. It remains as solid candidate to explain the chromospheric heating \citep{Jefferies+etal2006, Kalkofen2007, BelloGonzalez+etal2010b, Kanoh+etal2016, Grant+etal2018, Abbasvand+etal2020}, although some works have claimed that the acoustic wave flux is insufficient to heat quiet-Sun \citep{Fossum+Carlsson2005,Fossum+Carlsson2006, Carlsson+etal2007} and sunspot \citep{Felipe+etal2011} chromospheres. The determination of the propagating nature of compressive waves has fundamental implications to estimate their contribution to the heating. Recently, \citet{Rajaguru+etal2019} measured an acoustic energy flux in the 2-5 mHz frequency range between the upper photosphere and lower chromosphere of quiet-Sun regions larger than previous estimates. These results go in the line of those reported by \citet{Jefferies+etal2006}, who showed that inclined magnetic field can facilitate the upward propagation of low-frequency waves (below 5 mHz) through the ramp effect, providing a significant source of energy to the solar chromosphere. However, \citet{Rajaguru+etal2019} determined that the frequency of the waves that carry this additional energy flux is below the theoretical cutoff value, even if the effect of magnetic field inclination is taken into account. Our analyses can qualitatively explain the observational measurements from \citet{Rajaguru+etal2019} as a reduction of the cutoff produced by radiative losses. We have found that at heights between $z=200$ km and $z=400$ km, radiative losses can reduce the cutoff to values below 4 mHz, although \citet{Rajaguru+etal2019} detected propagation at even lower frequencies, in the range 2-4 mHz.    

Our results indicate that quiet-Sun regions with a stronger magnetic field can show a lower cutoff frequency (Fig. \ref{fig:cutoff_QS_Bz}), in agreement with \citet{Jefferies+etal2019}, although in our measurements the detailed variation of the cutoff with the field strength depends on the atmospheric height and stratification. In addition, our simulations prove the lower cutoff in regions with an inclined magnetic field (and $\beta<1$) due to the reduced gravity and the lower temperature gradients along the field lines (Fig. \ref{fig:cutoff_spot_incl}). The effect of the gravity was predicted by \citet{Bel+Leroy1977}. It has been confirmed through observations \citep{McIntosh+Jefferies2006,Jefferies+etal2006,Rajaguru+etal2019} and numerical simulations \citep{DePontieu+etal2004, Hansteen+etal2006, Heggland+etal2011}. To the best of our knowledge, the additional cutoff reduction produced by inclined magnetic fields in regions with significant gradients in the temperature is reported here for the first time. Although the reduced cutoff associated with magnetic fields in relatively quiet regions can be seen as a mechanism to supply additional acoustic flux to the chromosphere, several works have found that in the surrounding regions the upward energy flux is reduced \citep{Vecchio+etal2007, Rajaguru+etal2019}. \citet{Jefferies+etal2019} identified a larger cutoff around magnetic regions. This is consistent with the increase of the cutoff frequency as solar activity increases, as measured from low-degree modes \citep{Jimenez+etal2011}. The cutoff frequency is a fundamental parameter for helioseismology analyses since it determines the upper boundary of the p-mode resonant cavities. It has been found to be a major contributor to the travel-time shifts measured in sunspots using local helioseismic methods \citep{Lindsey+etal2010, Schunker+etal2013, Felipe+etal2017b}.

\section{Conclusions}
\label{sect:conclusions}

In this paper, we have examined the wave propagation between the solar photosphere and chromosphere using numerical simulations. We have focused on the evaluation of the cutoff frequency stratification, that is, the determination of the minimum frequency of propagating waves as a function of height in various solar models. The cutoff frequency was derived from the examination of phase difference spectra, by detecting the lowest frequency where a positive phase difference is measured. All these analyses have been performed for a set of standard solar atmospheric models representing quiet-Sun and umbral regions. We have evaluated several theoretical expressions commonly employed for deriving the cutoff. Our results show that, although the analytical cutoff frequencies exhibit a qualitative agreement with the actual values measured in the numerical simulations, none of them provides a notable match. The use of more refined cutoff formulae (\eg, Eqs. \ref{eq:wc2}-\ref{eq:wc4}) do not lead to significant improvements over the original cutoff expression for isothermal atmospheres (Eq. \ref{eq:wc1}). In fact, the latter expression, if applied accounting for its local variations in atmospheres with non-constant temperature, gives a better result in quiet-Sun regions with moderate magnetic field strength. The validity of the analytical expressions is also challenged by the assumptions employed in their derivation. Some of the most commonly used acoustic cutoff formulae neglect the effects of magnetic fields or radiative losses, which are fundamental to understand wave propagation in the solar atmosphere. We have found that radiative losses greatly reduce the photospheric cutoff frequency, and can partially explain some recent observations of propagation of low-frequency waves at photospheric heights in quiet Sun \citep{Wisniewska+etal2016, Rajaguru+etal2019} and sunspots \citep{Felipe+etal2018b}.

\begin{acknowledgements} 
Financial support from the State Research Agency (AEI) of the Spanish Ministry of Science, Innovation and Universities (MCIU) and the European Regional Development Fund (FEDER) under grant with reference PGC2018-097611-A-I00 is gratefully acknowledged. The authors wish to acknowledge the contribution of Teide High-Performance Computing facilities to the results of this research. TeideHPC facilities are provided by the Instituto Tecnol\'ogico y de Energ\'ias Renovables (ITER, SA). URL: \url{http://teidehpc.iter.es}.
\end{acknowledgements}

\bibliographystyle{aa} 
\bibliography{biblio.bib}

\end{document}